\input harvmac
\input amssym.def
\input epsf.tex
\def\N{{\cal N}}   
\def\E{{\cal E}}

\def\la{{\lambda}}  

\def\be{{\beta}}
\def\ga{\gamma}

\def\H{{\cal H}}

\def\g{{\frak g}}
\def\ie{{\it i.e.}}
\def\eg{{\it e.g.}}
\def\half{{1\over 2}}
\font\tenmsb=msbm10       \font\sevenmsb=msbm7
\font\fivemsb=msbm5       \newfam\msbfam
\textfont\msbfam=\tenmsb  \scriptfont\msbfam=\sevenmsb
\scriptscriptfont\msbfam=\fivemsb
\def\Bbb#1{{\fam\msbfam\relax#1}}

\def\Zop{{\Bbb Z}}

\def\bbbc{{\mathchoice {\setbox0=\hbox{$\displaystyle\rm C$}\hbox{\hbox
to0pt{\kern0.4\wd0\vrule height0.9\ht0\hss}\box0}}
{\setbox0=\hbox{$\textstyle\rm C$}\hbox{\hbox
to0pt{\kern0.4\wd0\vrule height0.9\ht0\hss}\box0}}
{\setbox0=\hbox{$\scriptstyle\rm C$}\hbox{\hbox
to0pt{\kern0.4\wd0\vrule height0.9\ht0\hss}\box0}}
{\setbox0=\hbox{$\scriptscriptstyle\rm C$}\hbox{\hbox
to0pt{\kern0.4\wd0\vrule height0.9\ht0\hss}\box0}}}}

\def\figin{\epsfcheck\figin}\def\figins{\epsfcheck\figins}
\def\epsfcheck{\ifx\epsfbox\UnDeFiNeD
\message{(NO epsf.tex, FIGURES WILL BE IGNORED)}
\gdef\figin##1{\vskip2in}\gdef\figins##1{\hskip.5in}
\else\message{(FIGURES WILL BE INCLUDED)}%
\gdef\figin##1{##1}\gdef\figins##1{##1}\fi}
\def\DefWarn#1{}
\def\figinsert{\goodbreak\midinsert}
\def\ifig#1#2#3{\DefWarn#1\xdef#1{fig.~\the\figno}
\writedef{#1\leftbracket fig.\noexpand~\the\figno}%
\figinsert\figin{\centerline{#3}}\medskip\centerline{\vbox{\baselineskip12pt
\advance\hsize by -1truein\noindent\footnotefont{\bf Fig.~\the\figno:} #2}}
\bigskip\endinsert\global\advance\figno by1}

\lref\gep{D. Gepner, {\it Fusion rings and geometry}, Commun. Math. Phys.
{\bf 141}, 381 (1991).} 

\lref\pz{V.B. Petkova, J.-B. Zuber, {\it Boundary conditions in charge
conjugate sl(N) WZW theories}, {\tt hep-th/0201239}.}
 
\lref\afflud{I. Affleck, A.W. Ludwig, {\it Universal noninteger `ground
state degeneracy' in critical quantum systems}, Phys. Rev. Lett.
{\bf 67}, 161 (1991).}
 
\lref\gg{M.R. Gaberdiel, T. Gannon, {\it Boundary states for WZW
models}, Nucl. Phys. {\bf B639}, 471 (2002); {\tt hep-th/0202067}.}
 
\lref\bou{P. Bouwknegt, P. Dawson, A. Ridout, {\it D-branes on group
manifolds and fusion rings}, JHEP {\bf 0212}, 065 (2002);
{\tt hep-th/0210302}.}
 
\lref\fs{S. Fredenhagen, V. Schomerus, {\it Branes on group manifolds,
gluon condensates, and twisted K-theory}, JHEP
{\bf 0104}, 007 (2001); {\tt hep-th/0012164}.}
 
\lref\mms{J. Maldacena, G. Moore, N. Seiberg, {\it D-brane instantons
and K-theory charges}, JHEP {\bf 0111}, 062 (2001);
{\tt hep-th/0108100}.}
 
\lref\braun{V. Braun, {\it Twisted K-theory of Lie groups},
{\tt hep-th/0305178}.}
 
\lref\mm{R. Minasian, G. Moore, {\it K-theory and Ramond-Ramond
charge}, JHEP {\bf 9711}, 002 (1997);
{\tt hep-th/9710230}.}
 
\lref\wittenK{E. Witten, {\it D-branes and K-theory}, JHEP
{\bf 9812}, 019 (1998); {\tt hep-th/9810188}.}
 
\lref\kapustin{A. Kapustin, {\it D-branes in a topologically
nontrivial B-field}, Adv. Theor. Math. Phys. {\bf 4}, 127 (2000);
{\tt hep-th/9909089}.}
 
\lref\boum{P. Bouwknegt, V. Mathai, {\it D-branes, B-fields and twisted
K-theory}, JHEP {\bf 0003}, 007 (2000); {\tt hep-th/0002023}.}
 
\lref\solit{J. Fuchs, C. Schweigert, {\it Solitonic sectors,
alpha-induction and symmetry breaking boundaries}, Phys. Lett.
{\bf B490}, 163 (2000); {\tt hep-th/0006181}.}
 
\lref\bfs{L. Birke, J. Fuchs, C. Schweigert, {\it Symmetry breaking 
boundary conditions and WZW  orbifolds}, Adv. Theor. Math. Phys.
{\bf 3}, 671 (1999); {\tt hep-th/9905038}.}
 
\lref\zuber{R.E. Behrend, P.A. Pearce, V.B. Petkova, J.-B. Zuber,
{\it Boundary conditions in rational conformal field theories},
Nucl. Phys. {\bf B579}, 707 (2000); {\tt hep-th/9908036}.}
 
\lref\kac{V.G. Kac, {\it Infinite dimensional Lie algebras}, Cambridge
University Press, Cambridge (1990) [3rd ed.].}
 
\lref\fss{J. Fuchs, B. Schellekens, C. Schweigert, {\it From Dynkin
diagram symmetries to fixed point structures},
Commun. Math. Phys. {\bf 180}, 39 (1996); {\tt hep-th/9506135}.}
 
\lref\go{P. Goddard, D.I. Olive, {\it Kac-Moody and Virasoro algebras
in relation to quantum physics}, Int. Journ. Mod. Phys. {\bf A1}, 303
(1986).}
 
\lref\gaberdiel{M.R. Gaberdiel, {\it Fusion of twisted
representations}, Int. Journ. Mod. Phys. {\bf A12}, 5183 (1997);
{\tt hep-th/9607036}.}
 
\lref\brb{N. Bourbaki, {\it Groupes et alg\`ebres de Lie, Chapitres
IV-VI}, Hermann, Paris (1968).}
 
\lref\dz{P. Di Francesco, J.-B. Zuber, {\it SU(N) lattice integrable
models associated with graphs}, Nucl. Phys. {\bf B338}, 602 (1990).}
 
\lref\qrs{T. Quella, I. Runkel, C. Schweigert, {\it An algorithm for
twisted fusion rules}, Adv. Theor. Math. Phys. {\bf 6}, 197 (2002); 
{\tt math.QA/0203133}.}
 
\lref\fsb{J. Fuchs, C. Schweigert, {\it Symmetries, Lie
Algebras, and Representations}, Cambridge University Press, Cambridge
(1997).} 
 
\lref\wlt{M.A. Walton, {\it Algorithm for WZW fusion rules: A proof},
Phys. Lett. {\bf B241}, 365 (1990).}

\lref\quella{T. Quella, {\it Branching rules of semi-simple Lie
algebras using affine extensions}, J. Phys. {\bf A35}, 3743 (2002);
{\tt math-ph/0111020}.}

\lref\gannon{T. Gannon, {\it Modular data: the algebraic
combinatorics of conformal field theory}, {\tt math.QA/0103044}.}

\lref\pss{J. Patera, R.T. Sharp, R. Slansky, 
{\it On a new relation between semisimple Lie algebras},
J. Math. Phys. {\bf 21}, 2335 (1980).}

\lref\be{R.E. Behrend, D.E. Evans, {\it Integrable Lattice Models for
Conjugate $A^{(1)}_n$}, {\tt hep-th/0309068}.}

\lref\afqs{A. Yu Alekseev, S. Fredenhagen, T. Quella, V. Schomerus,
{\it Non-commutative gauge theory of twisted D-branes}, Nucl. Phys. 
{\bf B646}, 127 (2002); {\tt hep-th/0205123}.}
 

\Title{\vbox{\baselineskip12pt
\hbox{hep-th/0311242}}}
{\vbox{\centerline{The charges of a twisted brane}}}
\smallskip
\centerline{Matthias R. Gaberdiel%
\footnote{$^\ast$}{{\tt mrg@phys.ethz.ch}}}
\smallskip
\centerline{\it Institute for Theoretical Physics, ETH H\"onggerberg}
\centerline{\it CH-8093 Z\"urich, Switzerland}
\bigskip
\centerline{and}
\bigskip
\centerline{Terry Gannon%
\footnote{$^\star$}{{\tt tgannon@math.ualberta.ca}}}
\smallskip
\centerline{\it Department of Mathematical Sciences, University of
Alberta}
\centerline{\it Edmonton, Alberta, Canada, T6G 2G1}\bigskip
\medskip
\vskip1.5cm
\centerline{\bf Abstract}
\bigskip
\noindent The charges of the twisted D-branes of certain WZW models are
determined. The twisted D-branes are labelled by twisted representations
of the affine algebra, and their charge is simply the ground state
multiplicity of the twisted representation. It is shown that the
resulting charge group is isomorphic to the charge group of the
untwisted branes, as had been anticipated from a K-theory
calculation. Our arguments rely on a number of non-trivial Lie
theoretic identities.  
\Date{November, 2003}

\newsec{Introduction}

The dynamics of D-branes in string theory is largely determined in
terms of their conserved charges. It is therefore of some significance
to determine these charges, as well as the resulting charge
groups. For strings propagating on the group manifold $G$ (for which
the world-sheet theory is described by a WZW model with symmetry
algebra $\g_k$), the brane charges can be determined directly in terms
of the underlying conformal field theory. Indeed, it was argued in
\refs{\fs} that for D-branes that preserve the affine symmetry (and
that are then labelled by the integrable highest weight
representations $P_+^k(\g)$), the charge $q_\mu$ of the brane labelled
by $\mu$ satisfies   
\eqn\chargedef{
\dim(\lambda)\, q_\mu =
\sum_{\nu\in  P_+^k(\g)} N_{\la \mu}{}^\nu \, q_\nu\qquad
({\rm mod}\ M)\,.} 
Here $\lambda\in P_+^k(\g)$ is an integrable highest weight
representation of $\g_k$, and $N_{\la \mu}{}^\nu$ are the fusion
rules. For finite $k$ this identity is only true modulo some integer
$M$, and the charge group that is carried by these D-branes is then 
$\Zop/M\Zop$, where $M$ is the largest positive integer for which 
\chargedef\ holds. [Since we can divide the $q_\mu$ by a common
factor, we are assuming here that the greatest common factor of the
$q_\mu$ is one.]
By taking $\mu$ to be the identity representation of $\g_k$, it follows
that 
\eqn\untwistedres{
q_\mu=\dim(\mu)\,.}
The number $M$ is then the largest number for which 
\eqn\untwischarg{
\dim(\lambda) \dim(\mu) =
\sum_{\nu\in  P_+^k(\g)} N_{\la \mu}{}^\nu \, \dim(\nu)\qquad
({\rm mod}\ M)}
holds. This number is always
\footnote{$^\star$}{In \refs{\bou}\ it is claimed there are some
exceptional levels, \eg\ $k=1$ for so$(n)$, where $M$ is larger than
(1.4) would indicate. However, K-theoretic considerations
\refs{\braun}\ suggest that $M$ is in fact always given by (1.4), 
even for small $k$, and that $M$ must also satisfy \untwischarg\ for
any {\it dominant} weight $\lambda$.} 
of the form
\eqn\nform{
M={k+h^\vee\over{\rm gcd}\{k+h^\vee,L\}}\,,}
where $h^\vee$ is the dual Coxeter number of $\bar{\g}$. It was shown
in \refs{\fs,\mms} that for $\bar{\g}=A_n$, for which $h^\vee=n+1$, $L$
is given by
\eqn\yar{
L_A={\rm lcm}\{1,2,\ldots,n\}\,.}
The value of $M$, \ie\ $L$, for the other affine algebras has been
determined in \refs{\bou}
\eqn\Mdef{\eqalign{
B_n: & \qquad  L_B = {\rm lcm} \{1,2,\ldots, 2n-1\} \cr
C_n: & \qquad  L_C = {\rm lcm} \{1,2,\ldots, n, 1,3,5,\ldots, 2n-1 \} \cr
D_n: & \qquad  L_D = {\rm lcm} \{1,2,\ldots, 2n-3\} \,.}}
The formula for $C_n$ was proven in \refs{\bou}; the formulae for
$B_n$ and $D_n$ (as well as the other algebras) were checked
numerically up to very high levels. For completeness we also mention
the dual Coxeter numbers for these cases,
$$
h^\vee(B_n) = 2n-1 \,, \qquad
h^\vee(C_n) = n+1 \,, \qquad
h^\vee(D_n) = 2n-2 \,.
$$

It has been known for some time that many WZW models also possess 
D-branes that only preserve the affine symmetry up to some
twist. In fact, for each automorphism $\omega$ of the corresponding
finite dimensional Lie algebra $\bar{\g}$, there exist
$\omega$-twisted D-branes. These D-branes are parametrised by
$\omega$-twisted highest weight representations of $\g_k$ 
\refs{\bfs,\solit,\gg}. It was argued in \refs{\fs} that the charge
group of the $\omega$-twisted D-branes of a WZW model is of the form
$\Zop_{M^{\omega}}$, where $M^\omega$ is characterised by the analogue
of \chargedef: the $\omega$-twisted D-brane labelled by $a$ has an
integer charge $q_a^\omega$, such that 
\eqn\twischargdiff{
\dim(\la) \, q_a^\omega = \sum_{b} {\cal N}_{\la a}{}^{b} \,
q_b^\omega \qquad({\rm mod}\ M^\omega)\,,}
where ${\cal N}_{\la a}{}^{b}$ are the corresponding NIM-rep
coefficients that appear in the Cardy analysis of these
branes. $M^{\omega}$ is then the largest integer for which
\twischargdiff\ holds, assuming again that the greatest common divisor
of the $q_a^\omega$ is one. Unlike the situation above, none of the
labels $a$ plays the role of an identity field, and thus it is not
clear {\it a priori} how to determine $q_a^\omega$, let alone
$M^\omega$. 

It was argued in \refs{\gg} (see also \refs{\solit}) that the NIM-rep
coefficients ${\cal N}_{\la a}{}^{b}$ can be identified with the
twisted fusion rules that describe the fusion of the twisted
representation $a$ with the untwisted representation $\lambda$ to give
the twisted representation $b$. The conformal highest weight spaces of
all three representations in question, $\la,a$ and $b$, form actually
representations of the {\it invariant subalgebra} $\bar\g^\omega$ that 
consists of the $\omega$-invariant elements of $\bar\g$. (This will
be discussed in detail in the following subsection.) As such, the
twisted fusion rules ${\cal N}_{\la a}{}^{b}$ must be truncations of
the tensor product coefficients of $\bar{\g}^\omega$. The situation is 
therefore analogous to the situation for the untwisted D-branes,
where the fusion rule coefficients are truncations of tensor product
coefficients of $\bar\g$. This suggests then that $q_a^\omega$ should 
simply be the dimension of the conformal highest weight space of the  
$\omega$-twisted highest weight representation
$a$, \ie\ the Weyl dimension of the representation of the
corresponding invariant subalgebra
$\bar{\g}^\omega$\footnote{$^\dagger$}{A related proposal was put
forward in \refs{\afqs}, based on an analysis for large
level. However, the proposal of \refs{\afqs} differs from ours in that
for su$(2n+1)$, they take the relevant `invariant subalgebra' to be
so$(2n+1)$, whereas it is sp$(2n)$ in our case.}  
\eqn\ansatz{q_a^\omega={\rm dim}_{\bar{\g}^\omega}(a)\,.}
In this paper we shall show that, up to an appropriate equivalence,
this is the unique solution of \twischargdiff, and that it solves
\twischargdiff\ with $M^\omega=M$. We shall explain this construction
in detail for the case of the classical algebras with $\omega$ being
charge conjugation (or chirality flip for the case of
so$(2n)$). Looked at from a purely Lie theoretic perspective, the
resulting set of identities is quite remarkable. The generalisation to
the two other cases (involving $D_4$ and $E_6$) should be
straightforward. In the course of the proof we shall also give very
simple and explicit formulae for the corresponding NIM-rep graphs,
from which it will be manifest that they are truncations of the
tensor product coefficients of $\bar{\g}^\omega$. This lends strong
support to the suggestion  that they are in fact the twisted fusion
rules. Other manifestly integral formulae for these coefficients are
given in \refs{\qrs}.  

It is believed that the D-brane charge groups coincide with certain 
untwisted and twisted geometric K-theory groups
\refs{\mm,\wittenK,\kapustin,\boum}. The latter have been 
recently calculated \refs{\braun}, and it was found that
$M^\omega=M$. Our conformal field theory calculation is therefore in
nice agreement with this result.
\medskip
 
The paper is organised as follows. In the remainder of this section we
collect some general observations about the structure of the twisted
algebras, motivate our ansatz, and summarise our results. In section~2,
the case of ${\rm su}(3)$ is analysed in detail. This is generalised
to ${\rm su}(2n+1)$ in section~3. Section~4 deals with the analysis
for ${\rm su}(2n)$, and the case of ${\rm so}(2n)$ is described in 
section~5. In section~6 we show that these solutions are in fact
unique, and section~7 contains our conclusions. There are a number of
appendices in which the details of some of the more Lie theoretic
calculations have been collected.  
 
 
 
\subsec{Background information}
 
We assume the reader is familiar with the basics of non-twisted affine
algebras (for this see \eg\ \refs{\kac,\fsb}). 
Let $\g_k$ be any affine (non-twisted) algebra, with colabels
$a^\vee_i$, $i=0,\ldots, n$. Then the dual Coxeter number $h^\vee$
equals $\sum_{i=0}^n a_i^\vee$, and the level $k$ weights 
$\la\in P_+^k(\g)$ are all 
$(\la_0;\la_1,\ldots,\la_n) =\sum_{i=0}^n\la_i\Lambda_i$,
which are dominant (i.e.\ have $\la_i\ge 0$, $\la_i\in\Zop$), and 
obey $\sum_{i=0}^n \la_i a_i^\vee=k$.
We will usually identify the affine weight $(\la_0;\la_1,\ldots,\la_n)$
with the finite weight $[\la_1,\ldots,\la_n]$.

Suppose now that $\omega$ is an outer automorphism of the underlying
finite dimensional Lie algebra $\bar{\g}$. A representation $a$ of the 
twisted algebra $\g^\omega$ is the same as a twisted representation of
the algebra $\g$ (see for example \refs{\go} for an introduction into
these  matters). If we choose a basis for the finite dimensional Lie
algebra $\bar{\g}$ to consist of eigenvectors of the automorphism
$\omega$, then this is a representation for which the modes $J^a_n$
for which $a$ has $\omega$-eigenvalue $+1$ have $n\in\Zop$, while the
modes $J^b_r$ for which $b$ has $\omega$-eigenvalue $-1$ have
$r\in\Zop+\half$. [We are assuming here for simplicity that $\omega$
is of order two, as will be the case for the examples discussed in
this paper.] Let us denote by $\bar{\g}^\omega$ the {\it invariant
subalgebra} of $\bar{\g}$, \ie\ the subalgebra that consists of the
$\omega$-invariant generators. The generators in $\bar{\g}^\omega$
commute with $L_0$, and they act on the space of conformal highest
weight states of $a$. In fact, if $a$ is an irreducible representation
of $\g^\omega$, the conformal highest weight states will form an
irreducible representation of $\bar{\g}^\omega$, and the
representation $a$ is uniquely characterised in terms of this
representation. Thus we can think of the labels that describe the
integrable $\omega$-twisted representations of $\g_k$ as describing
representations of $\bar{\g}^\omega$.  
\bigskip

\noindent {\it 1.1.1. Untwisted and twisted D-branes}
\smallskip

In this paper we shall consider strings propagating on the
simply-connected group $G$, for which the corresponding conformal
field theory is just the charge conjugation modular invariant
associated to $\g_k$. In this case, the untwisted D-branes are 
labelled by the highest weight representations in $P_+^k(\g)$, and are
simply given by the Cardy formula
\eqn\cardy{
|\!| \lambda \rangle\!\rangle = \sum_{\mu\in P^k_+} 
{S_{\lambda \mu} \over \sqrt{S_{0\mu}}} \, | \mu\rangle\!\rangle\,,}
where $| \mu\rangle\!\rangle$ is the Ishibashi state in the
representation $\H_\mu\otimes \bar\H_\mu^\ast$, and $S_{\lambda\mu}$
is the modular $S$-matrix. (For an introduction into these matters see
for example \refs{\zuber}.) The open string that stretches between the
branes labelled by $\lambda$ and $\mu$ then contains the
representation $\H_\nu$ of $\g_k$ with multiplicity
\eqn\verlinde{
N_{\nu \mu}{}^{\lambda} = \sum_{\rho\in P^k_+} 
{S_{\mu\rho} \, S_{\nu\rho} \, S_{\lambda \rho}^\ast \over S_{0\rho}}
\,.}
Because of the Verlinde formula, these numbers are just the fusion
rules of $\g_k$.

Suppose now that $\omega$ is an outer automorphism of the finite
dimensional Lie algebra $\bar\g$. Then $\omega$-twisted Ishibashi
states exist for the subset ${\cal E}$ of {\it exponents}: this
consists simply of the $\omega$-invariant representations of $P^k_+$. 
It was argued in \refs{\gg} (see also \refs{\bfs,\solit,\pz}) that 
the $\omega$-twisted D-branes are then labelled by the integrable
$\omega$-twisted representations $a$ of $\g_k$, and that they are
explicitly given as 
\eqn\twiscardy{
|\!| a \rangle\!\rangle^\omega = \sum_{\mu\in \E} 
{\psi_{a \mu} \over \sqrt{S_{0\mu}}} \, | \mu\rangle\!\rangle^\omega\,,}
where $\psi$ describes the modular transformation of twisted and
twining characters. In order to describe this in more detail, we
define, for each $\mu\in\E$, the {\it twining character} 
\eqn\twining{
\chi_\mu^{(\omega)}(\tau) =
{\rm Tr}_{\H_\mu}\left(\tau_\omega q^{L_0-{c\over 24}}\right)\,,}
where $\tau_\omega$ is the induced action of $\omega$ on $\H_\mu$, the
representation space corresponding to $\mu$. An important and
nontrivial fact is that the twining character agrees precisely with
the ordinary character of the so-called orbit Lie algebra $\check{\g}$
\refs{\fss}. Upon a modular $S$-transformation we then have
\eqn\modtwis{
\chi_{\mu}^{(\omega)}(-1/\tau) = \sum_{a}
\psi_{a \mu}\, \chi_{a}(\tau)\,,}
where the $\chi_{a}$ are characters of the twisted algebra
$\g^\omega$, and the modular matrix $\psi_{a\mu}$ is the matrix that
appears in \twiscardy. The open string that stretches between the
branes labelled by $b$ and $a$ then contains the
representation $\H_\nu$ of $\g_k$ with multiplicity
\eqn\twisverlinde{
{\cal N}_{\nu a}{}^{b} = \sum_{\rho\in \E} 
{\psi_{a\rho} \, S_{\nu\rho} \, \psi_{b \rho}^\ast \over S_{0\rho}}
\,.}
The consistency of the construction requires that these numbers are
non-negative integers, and in fact, they must define a {\it NIM-rep}
of the fusion algebra -- see \eg\ \refs{\zuber,\gannon} for an
introduction to these matters. The positivity of these numbers at
least for $A_n$ is a consequence of the subfactor realisation of this
NIM-rep \refs{\be}. It was argued in \refs{\gg} (see also
\refs{\solit}) that these numbers agree in fact with the twisted
fusion rules. (The fusion of twisted representations is for example
discussed in \refs{\gaberdiel}.) At any rate, these are the numbers
that appear in the definition of the twisted charges \twischargdiff.  
\smallskip

Explicit formulae for the $\psi$-matrix, as well as for the NIM-reps 
${\cal N}_{\lambda a}{}^b$, were given in \refs{\gg}. Actually, 
a dramatic simplification occurs if we restrict in the NIM-rep
formulae $\lambda$ to the fundamental weights $\la=\Lambda_i$ (see
also \refs{\pz}). This does not lose any information since the
character ring of $\bar{\g}$ is generated by the characters of the 
$\Lambda_i$, and we can therefore write {\it any} NIM-rep coefficient
${\cal N}_{\la a}{}^b$ in terms of the various coefficients
${\cal N}_{\Lambda_i a'}{}^{b'}$.  
More precisely, write the $\bar{\g}$-character ch$_\la$ as a
polynomial $P_\la({\rm ch}_{\Lambda_1},{\rm ch}_{\Lambda_2},\ldots)$
in the fundamentals; then the NIM-rep matrix ${\cal N}_\la$ 
equals $P_\la({\cal N}_{\Lambda_1},{\cal N}_{\Lambda_2},\ldots)$. For
all of the NIM-reps considered in this paper, we will explicitly write
simple expressions for the ${\cal N}_{\Lambda_i a}{}^b$.

As an aside, note that the limit of \modtwis\ as
$\tau\rightarrow i\infty$ (\ie\ $q\rightarrow 0$) will be dominated by
the contribution from the unique weight of $\g^\omega$, namely
$a=0$, with minimal conformal weight. The dominant term on the right
hand side of \modtwis\ in this limit is $\psi_{0\mu} q^{-c/24}$.
If we take the $\tau$ limit along the imaginary axis, the left-side is
manifestly positive (being an ordinary character of the orbit Lie
algebra). Thus we obtain that $\psi_{0\mu}\ge 0$ for all
$\mu$. Similarly, $\psi_{a0}\ge 0$ for all $a$. Both of these are
consistent with the observations of \refs{\gg}; the positivity of
$\psi_{a0}$ is also required by its interpretation in \afflud\ as a
boundary entropy. 
\bigskip

\noindent {\it 1.1.2. More details about the relevant twisted algebras}
\smallskip

In this paper we shall only deal with the classical algebras that have
a non-trivial order-2 automorphism (\eg\ charge-conjugation). The
relevant algebras $\bar{\g}$ are then ${\rm su}(n)$, and 
${\rm so}(2n)$. The invariant subalgebra $\bar{\g}^\omega$
of ${\rm su}(2n+1)$ and ${\rm su}(2n)$ is $C_n$, while the invariant
subalgebra of ${\rm so}(2n)$ is $B_n$. This can be read off from the
structure of the weights of the twisted algebra. For example, the data
on page~81 of \refs{\kac} can be used to compute \eg\ the inner
products $(\Lambda_i|\Lambda_j)$ of fundamental weights (these are
defined to be dual to the coroots) and hence identify
$a=\sum a_i\Lambda_i$ {\it etc}. 

Similarly, one can also identify the
labels of the $\omega$-invariant weights $\mu$ with weights of some
finite dimensional subalgebra. For the case of su$(2n+1)$, $\mu$ can
be identified with a weight of $C_n$, while for su$(2n)$, $\mu$ is
a weight of $B_n$, and for so$(2n)$, $\mu$ is a weight of $C_n$. [Another
way to make these identifications is to compute the conformal weights
of the exponents $\mu$, \ie\ to compute the norms
$(\mu+\rho| \mu+\rho)$ of the weights $C\mu=\mu$, and to compare them
with the conformal weights in $C_n$ and $B_n$, respectively.]

Thus the labels $\mu$ and $a$ appear symmetrically for
su$(2n+1)$, while they are genuinely different for su$(2n)$ and
so$(2n)$. This can be traced back to the fact that both the rows
and columns of the $\psi$ matrix for su$(2n+1)$ label representations
of $A_{2n}^{(2)}$. On the other hand, for ${\rm su}(2n)$ the rows
label representations of $A_{2n-1}^{(2)}$ while the columns label
representations of $D_{n+1}^{(2)}$, and for ${\rm so}(2n)$ the rows
label representations of $D_{n}^{(2)}$ while the columns label
representations of $A_{2n-3}^{(2)}$.

Finally, it remains to determine the levels at which these invariant
subalgebras appear. The easiest way to anticipate the relevant shift
in level, is to note that all the relevant
formulae (\eg\ the modular matrices) depend directly on
$\kappa=k+h^\vee$ rather than $k$ itself. Thus if we  convert say
su$(2n+1)$ data (where $\kappa=k+2n+1$) into so$(2n+1)$ data (where
$\kappa=k_B+2n-1$), we should shift the effective level to $k_B=k+2$.

\subsec{Summary of our results}

Let us state clearly our results and assumptions. There are three
conjectures that are required for some parts of our arguments. These
are:
\medskip

\noindent{{\bf Conjecture B:}} The largest number $m$ (call it $m_B$) 
such that 
\eqn\conj{{\rm dim}(\la)\,{\rm dim}(\mu)=
\sum_\nu N_{\la\mu}^\nu{\rm dim}(\nu)\qquad ({\rm mod}\ m)} 
for all $B_n$ dominant weights $\la$ and all $\mu\in P_+^k(B_n)$, is 
$M_B=(k+2n-1)/{\rm gcd}\{k+2n-1,L_B\}$ for $L_B$ given in \Mdef.
\medskip

\noindent{{\bf Conjecture D:}} The largest number $m$ (call it $m_D$) 
such that \conj\ holds for all $D_n$ dominant weights $\la$ and 
all $\mu\in P_+^k(D_n)$, is 
$M_D=(k+2n-2)/{\rm gcd}\{k+2n-2,L_D\}$ for $L_D$ given in \Mdef. 
\medskip

\noindent{{\bf Conjecture B$^{spin}$:}} The largest number $m$  
(call it $m_B^{spin}$) such that \conj\ holds for all $B_n$ non-spinor
dominant weights $\la$ and all spinors $\mu\in P_+^k(B_n)$ (\ie\ 
$\mu_n$ is odd and $\la_n$ is even), is $2^nM_B$.
\medskip

Conjectures B and D were made in \refs{\bou} and are (to our knowledge)
still unproven, although the analogues for $A_n$ and $C_n$ have been
proven. There is considerable numerical support for their general
validity, and the fact that they fit so nicely into our context lends
further support. Conjecture B$^{spin}$ is a natural generalisation of
Conjecture B. In appendix~D we prove it (\ie\ reduce it to Conjecture
B) whenever 4 does not divide $M_B$. We also show $m_B^{spin}$ must 
equal some power of 2 times $M_B$, and must divide $2^nM_B$. In
\refs{\bou}\ a generating set for each fusion ideal ${\cal I}_k$ is
conjectured; that conjecture implies  
\eqn\conjtwo{ m_B^{spin}={\rm gcd}\{{\rm
dim}_B((k-2)\Lambda_1+\Lambda_i+\Lambda_n)\}_{1<i<n}\,.}
\bigskip

\noindent Our assertion about the twisted charge group consists of
two separate statements:
\medskip

\noindent (i) eq.\ \ansatz\ satisfies \twischargdiff\ with
$M^\omega=M$; 

\noindent (ii) eq.\ \ansatz\ is the unique solution to \twischargdiff,
up to rescaling. (Of course, given any solution $q_a^\omega$ to
\twischargdiff, another will be given by $q_a^\omega{}':=c\, q_a^\omega$
for any number $c$. What (ii) says is that any solution to
\twischargdiff\ is of that form, for $q_a^\omega$ given by \ansatz.) 
\bigskip

We should note that $M^\omega$ always has to be a factor of 
$M$.\footnote{$^\ddagger$}{The following argument is due to Stefan
Fredenhagen; we are grateful to him for communicating this argument to
us.} By construction, $M$ is the greatest common divisor of the
dimensions of elements in the fusion ideal, the ideal by which we
have to quotient the representation ring in order to obtain the fusion
ring. Since the NIM-rep is a representation of the fusion ring, any
element of the fusion ideal must act trivially. Thus for any
$\alpha$ in the fusion ideal and for any twisted D-brane $a$, 
$\dim(\alpha)\, q_a^{\omega} = 0 \, ({\rm mod}\ M^\omega)$. Together
with the fact that the $q_a^{\omega}$ must be coprime, this implies
that $M^\omega$ must be a factor of $M$. 
\smallskip

For $\bar{\g}={\rm su}(2n+1)$, when the level $k$ is odd, we {prove}
(i) and (ii), with no additional assumptions.

For $\bar{\g}={\rm su}(2n+1)$, when the level $k$ is even, we require
Conjecture B. More precisely, we only need that conjecture when we
state that the twisted charge group $M^\omega$, which we show equals
$m_B$, in fact equals the untwisted $M_A$. 

For $\bar{\g}={\rm su}(2n)$, when the level is odd, we again only
require Conjecture B. For $\bar{\g}={\rm su}(2n)$, when the level is
even, we require in addition Conjecture B$^{spin}$ for the proof of
(i).

For $\bar{\g}={\rm so}(2n)$, (i) follows from Conjectures B
and D only, unless $2^{n+1}$ divides $M_B$, in which case Conjecture
B$^{spin}$ is also needed.  Only Conjectures B and D are needed for 
(ii).

\newsec{The construction for ${\rm su}(3)$}
 
Let us first illustrate the construction in the simplest non-trivial
case, where $\bar{\g}={\rm su}(3)$. As follows from \nform, the untwisted
charge lattice is $\Zop/M \Zop$, where $M=k+3$ if $k$ is even, and
$M=(k+3)/2$ if $k$ is odd. The NIM-rep matrices
${\cal N}_{\Lambda_i a}{}^b$ involving both fundamental weights
$\Lambda_1=\,${\bf 3} and $\Lambda_2={\bf \bar{3}}$ are described by
the following `twisted fusion graph' of \refs{\dz,\gg}. [From now on 
we shall call this NIM-rep the twisted fusion rules; our
analysis does, however, not rely on the assumption that this
identification is in fact correct.]
\ifig\sunpic{The NIM-rep graphs for su$(3)$ with charge conjugation
for $k=1,2,3,4,5$.}
{\epsfxsize4.5in\hskip.2cm\epsfbox{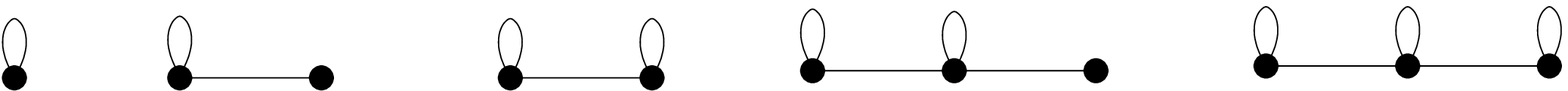}}

Let us first consider the case when $k$ is even. There are $k/2+1$
twisted representations (corresponding to each of the nodes on the fusion
graph), which we may label left to right by $a=0,1,\ldots,k/2$. The
twisted fusion rules are uniquely determined in terms of the fusion
with the fundamental representation ${\bf 3}$ of ${\rm su}(3)$, and
are given as 
\eqn\suthree{\eqalign{
{\bf 3} \otimes [0] & = [0] + [1] \,, \cr
{\bf 3} \otimes [n] & = [n-1] + [n] + [n+1]\,,
 \qquad n=1,\ldots,k/2-1 \,, \cr
{\bf 3} \otimes [k/2] & = [k/2-1] \,.
}}
These fusion rules imply that the corresponding charges
$q_{a}\equiv q_a^\omega$, where $\omega$ is charge conjugation, must
satisfy  
\eqn\suthreeo{\eqalign{
3 \, q_0 & = q_0 + q_1  \cr
3 \, q_{n} & = q_{n-1} + q_n + q_{n+1}
\qquad n=1,\ldots,k/2-1 \,, \cr
3 \, q_{k/2} & = q_{k/2-1} \,,}}
where all equalities are understood to be modulo some as-yet-undetermined
number $M^\omega$. Starting with the initial condition $q_0=1$ say,
the unique solution (mod $M^\omega$) to these recursion relations is
\eqn\suthreea{
q_n = n+1 \,.}
In the next section we will interpret this as a dimension. These
values satisfy all but the last equation identically; the last 
equation then gives
\eqn\suthreee{
3 (k/2+1) = k/2 \qquad ({\rm mod}\ M^\omega)\,,}
and thus we should take $M^\omega=k+3$, in agreement with the value \nform\
for $M$. If we had instead performed the analysis starting with the 
`initial condition' ${q}_{k/2}=1$ (which would have been natural had
we numbered the nodes of Figure 1 from right to left), we would have
obtained the unique answer 
\eqn\suthreeg{
\hat{q}_n = 2 (k/2-n)+1 \,,}
again with $M^\omega=k+3$. This solution is equivalent to \suthreea\
since
\eqn\suthreeh{
\hat{q}_n  = (k+3) - 2 (n+1) = M^\omega - 2 q_n \,.}
Indeed, since the charges $q_n$ are only defined
modulo $M^\omega$, the two solutions differ by multiplication by
$l=-2$. Since $l=-2$ is coprime to $M^\omega=k+3$ here, the two solutions
are mathematically interchangeable. More generally, we call two sets
of charges $q_a^\omega$ and $\hat{q}_a^\omega$ {\it equivalent} whenever
\eqn\lequiv{
\hat{q}_a^\omega = l\, q_a^\omega \qquad ({\rm mod}\ M^\omega) \,,
}
for some $l\in\Zop$ coprime to $M^\omega$. This is the obvious
redundancy in the solutions to \suthreeo. Nevertheless, `equivalent'
solutions are not equally satisfactory: \suthreea\ has a direct
interpretation in terms of dimensions which is much more obscure in
\suthreeg. Starting from the fusion graphs of Fig.~1, it is unclear
which of these mathematically equivalent solutions is physically
preferred. However, from the analysis of \refs{\gg} it is clear that
the left-most vertex (what we called $a=0$) corresponds to the
identity and should be assigned charge $q_0=1$. We will  
see in the next section that the first equation 
${\bf 3}\otimes[0]=[0]+[1]$ then corresponds to the 
su$(3)\subset$ su$(2)$ branching rule ${\bf 3}={\bf 1}+ {\bf 2}$. 
\medskip

The analysis for $k$ odd is essentially the same. For $k$ odd, there
are $(k+1)/2$ twisted representations that we label by the integers
$a_1=0,1,\ldots,(k-1)/2$. The twisted fusions with ${\bf 3}$ are the
same as for even $k$, except that now the last fusion is
\eqn\suthreeodd{
{\bf 3} \otimes [(k-1)/2]  = [(k-3)/2] + [(k-1)/2] \,.
}
The recursion formula still implies that the charges are given by
\suthreea, but now the last equation gives
\eqn\suthreeb{
(k+1) = (k-1)/2 \qquad ({\rm mod}\ M^\omega) \,,}
which therefore implies that the maximal choice for $M^\omega$ is
$M^\omega=(k+3)/2$, and thus agrees with the untwisted $M$.

\newsec{The analysis for su$(2n+1)$}

Let us now generalise the previous discussion to $A_{2n}$ with
$\omega=C$ being charge conjugation. As was explained in
\refs{\bfs,\gg}, the $\omega$-twisted boundary states are 
labelled by the level $k$ $\omega$-twisted weights of
$\g=\widehat{{\rm su}}(2n+1$), or alternatively the level $k$ weights
of the twisted Lie algebra $\g^\omega=A_{2n}^{(2)}$. They can be
equated with all ($n+1$)-tuples $(a_0;a_1,\ldots,a_n)$ where
$k=a_0+2a_1+2a_2+\cdots+2a_n$ and $a_i\in \Zop_{\ge 0}$. The ground
states of this twisted representation (\ie\ the states of lowest
conformal weight) form an irreducible representation of the invariant
subalgebra $\bar{\g}^\omega=C_n$ with highest weight
$[a_1,\ldots,a_n]$. We now propose that the corresponding D-brane
charge is simply the Weyl dimension of this irreducible
representation, \ie
\eqn\suoddp{
q_a^\omega = \dim_{C} \left( [a_1,\ldots,a_n] \right)=\dim_C(a) \,. }
In this section we shall prove, assuming conjecture B for $k$ even, 
that this solves \twischargdiff\ with $M^\omega=M$. In section~6 we
address uniqueness, and analyse whether $M^\omega$ can be
increased. For $n=1$, the invariant subalgebra is $C_1={\rm su(2)}$,
and the right hand side is just $a_1+1$, in agreement with \suthreea\
of the previous section.  

Our proof that \suoddp\ solves \twischargdiff\  depends on whether $k$
is even or odd, and we will deal with these cases in turn.

\subsec{The case of odd level}

When the level $k$ is odd, a simplification occurs in that we can
identify (bijectively) each boundary label $a$ with the $\widehat{C}_n$
level ${k-1\over 2}$ weight 
$\widetilde{a}=({a_0-1\over 2};a_{1},\ldots,a_n)\equiv [a_1,\ldots,a_n]$,
and similarly for $\mu$.\footnote{$^\star$}{The integrable highest
weights of $\widehat{C}_n$ level $\tilde{k}$ are the weights 
$\tilde{\mu}=(\tilde{\mu}_0;\tilde{\mu}_1,\ldots,\tilde{\mu}_n)$ for
which $\sum_{i=0}^{n} \tilde{\mu}_i=\tilde{k}$.}
(For convenience, we shall use tildes to denote quantities associated 
with $\widehat{C}_n$ level ${k-1\over 2}$.) With this, the NIM-rep
becomes \refs{\gg} 
\eqn\nimsuodkev{
\N_{\la a}{}^b=\sum_{\tilde{\ga}}b^\la{}_{\tilde{\ga}}\,
\widetilde{N}_{\tilde{\ga}\tilde{a}}{}^{\tilde{b}}\,,}
where $\oplus \,b^\la{}_{\tilde{\ga}}\,(\tilde{\ga})=(\la)$ are the
branching rules for $C_n\subset A_{2n}$. The fusion
coefficients $\widetilde{N}_{\tilde{\ga}\tilde{a}}{}^{\tilde{b}}$
are defined in the obvious way for $\tilde{\ga}$ not necessarily in
$\tilde{P}_+=P_+^{(k-1)/2}(C_n)$ (using \eg\ the fact that
$S_{\la\mu}$ is affine Weyl antisymmetric in $\la$ and $\mu$), in
which case they can be 
negative. For the fundamental representations $(\Lambda_i)$ of
$A_{2n}$, the branching rules are particularly simple:
\eqn\branchac{
(\Lambda_i) = (\Lambda_{2n+1-i})=[0,\ldots,0] + \bigoplus_{j=1}^{i}
\,\tilde{\Lambda}_j \qquad i\leq n
\,,}
where $\tilde{\Lambda}_j$ is the $j$th fundamental representation of
$C_n$. (Of course, the left hand side of \branchac\ refers to the
restriction of the $A_{2n}$ representations to the $C_n$
subalgebra.) As a consequence of \branchac, the corresponding NIM-reps 
are simply 
\eqn\simpleNIM{
\N_{\Lambda_i a}{}^b=\N_{\Lambda_{2n+1-i}a}{}^b= \delta_a{}^b
+ \sum_{j=1}^i \tilde{N}_{\tilde{\Lambda}_j \tilde{a}}{}^{\tilde{b}}
\qquad i\leq n \,.}
This formula was already found in \refs{\pz}. For completeness we 
mention that the NIM-rep coefficients for the simple currents $J^i$
are all trivial: $\N_{J^i\ a}{}^b=\delta_a{}^b$.   

As an aside, we note that it is manifest  from \nimsuodkev\ that the
NIM-rep coefficient $\N_{\la a}{}^b$ equals (for sufficiently large $k$)
the tensor product coefficient $T_{\la a}{}^b$ for the invariant
subalgebra $\bar{\g}^\omega=C_n$. This lends strong support to the
assertion that these NIM-rep coefficients are in fact the twisted
fusion rules. 

With the ansatz \suoddp\ for $q_a^\omega$, we can now rewrite the
right hand side of \twischargdiff\ as
\eqn\suoddkodda{\eqalign{
\sum_b \N_{\lambda a}{}^b q_b^\omega & =
\sum_{\tilde{\ga}} b^\la{}_{\tilde{\ga}}\,
\sum_{\tilde{b}} \widetilde{N}_{\tilde{\ga}\tilde{a}}{}^{\tilde{b}}
\dim_{C}(\tilde{b}) \cr
& = \sum_{\tilde{\ga}}b^\la{}_{\tilde{\ga}}\, \dim_{C}(\tilde{\ga})\,
\dim_{C}(\tilde{a}) \qquad ({\rm mod}\ M_C) \cr
& = \dim_{A} (\lambda) \, \dim_{C}(\tilde{a})\,
\qquad ({\rm mod}\ M_C) \,,}}
where $M_C$ is the untwisted charge number corresponding to $C_n$ at
level $(k-1)/2$ (see \Mdef), which was proved in \refs{\bou} to be
\eqn\suoddkoddb{
M_C = {(k-1)/2+n+1\over {\rm gcd}\{(k-1)/2+n+1,L_C\}} \,,}
where $L_C={\rm lcm}\{1,2,\ldots,n,1,3,\ldots,2n-1\}
={\rm lcm}\{1,2,\ldots,2n\}/2$.
Thus the denominator (like the numerator) of \suoddkoddb\ is half that of
\nform\ for $A_{2n}$, and it follows that $M_C$ agrees with the
untwisted $M$ for $A_{2n}$ at odd level $k$.

\subsec{The case of even level}

If $k$ is even, the previous analysis is not available, but one can
still regard the $\psi$-matrix as a symmetric submatrix of the
$S$-matrix for, in this case, $\widehat{B}_n$ level $k+2$.
More specifically, identify the boundary state $a$ with
$a'=(a_0+a_1+1;a_1,\ldots,a_{n-1},2a_n+1)\in P_+'\equiv 
P_+^{k+2}(B_n)$.\footnote{$^\dagger$}{$P_+^{k'}(B_n)$ consists of the
weights $\mu'=(\mu_0';\mu_1',\ldots,\mu_n')$ for which 
$\mu'_0+\mu'_1+2\sum_{i=2}^{n-1} \mu'_i+\mu'_n=k'$.}
(For convenience we shall use primes to denote quantities associated
with $\widehat{B}_n$ level $k+2$.) Using this identification, we can
express \refs{\gg}\ the NIM-rep coefficients
$\N_{\la a}{}^b$ in terms of
the ordinary fusion coefficients $N'$ of $\widehat{B}_n$ level $k+2$,
and branching rules for $B_n\subset A_{2n}$, namely
\eqn\nimsuodd{
\N_{\la a}{}^b=\sum_{\ga'}b^\la{}_{\ga'}\,
\bigl(N'_{\ga' a'} {}^{b'}-N'_{\ga' a'}{}^{J'b'}\bigr)\,.}
Here, $J'$ is the simple current of $\widehat{B}_n$, which acts on
weights $\nu'$ by $J'\nu'=(\nu'_1;\nu'_0,\nu'_2,\ldots,\nu'_n)$. The
coefficients $b^\la{}_{\ga'}$ describe the branching rules
$\oplus_{\ga'}\,b^\la_{\ga'}\,(\ga')=(\la)$ for the embedding of
$B_n\subset A_{2n}$. Note that these $\gamma'$ necessarily will be
non-spinors.
The sum in \nimsuodd\ is over all dominant weights $\ga'$ of $B_n$;
as before we extend the definition of the fusion coefficients
$N'_{\ga'a'}{}^{b'}$ to arbitrary dominant weights $\ga'$ in the
obvious way. Again, for the fundamental representations of $A_{2n}$,
the branching rules are very simple
\eqn\branchba{\eqalign{
(\Lambda_i)=(\Lambda_{2n+1-i}) & = (\Lambda_i') \qquad i<n \cr
(\Lambda_n)=(\Lambda_{n+1}) & = (2 \Lambda'_n) \,.}}
Thus, as before, \nimsuodd\ gives a very simple and explicit formula
for the corresponding NIM-reps:
\eqn\suodev{\N_{\Lambda_i a}{}^b=\N_{\Lambda_{2n+1-i} a}{}^b=
N'_{\Lambda'_i a'}{}^{b'}\,,}
with the obvious analogue for $\N_{\Lambda_n}$. These formulae were
already found in \refs{\pz}. For completeness we 
also mention that the NIM-rep coefficients of the simple currents are
trivial: ${\cal N}_{J^i\ a}{}^b=\delta_a{}^b$. 

For su(3), the above  requires some clarification.
By `$B_1$ level $k+2$' here we mean su(2) at level $2(k+2)$; 
$(a_0;a_1)'= (2a_0+2a_1+3;2a_1+1)$ and
$(\mu_0;\mu_1)'=(2\mu_0+2\mu_1+3;2\mu_1+1)$. The simple current acts
as $J'(\nu_0';\nu_1')=(\nu_1';\nu_0')$; \branchba\ becomes
$(\Lambda_1)=(\Lambda_2)=(2\Lambda_1')$. 

As an aside we note that the NIM-rep coefficient $\N_{\la a}{}^b$ in  
\nimsuodd\ does indeed equal the tensor product coefficient 
$T_{\la a}{}^b$ for  the invariant subalgebra $\bar{\g}^\omega=C_n$,
for sufficiently large level, as should be the case if the NIM-rep
agrees with the twisted fusion rules; this is illustrated for the case
of su$(7)$ in appendix~C. The proof of this not-completely-obvious
fact follows the analogous argument sketched in the next section for
su($2n$). Incidentally, the large $k$ limit of \nimsuodd\ tells us
that the $C_n$ tensor product coefficient $T_{\la a}{}^b$ equals the
$B_n$ tensor product coefficient $T'_{\la a'}{}^{b'}$ -- something
which is not {\it a priori} obvious.

Let us first make the ansatz (as we shall see later on this will turn
out to be equivalent to \suoddp)\
\eqn\suoddodda{
\hat{q}^\omega_{a} =
\dim_{B} \left( [a_1,\ldots,a_{n-1},2a_n+1] \right)=\dim_{B}(a') \,,} 
where the right hand side is the $B_n$ dimension of $a'$.
Let ${\cal G}$ be the subset
of $P_+^{k+2}(B_n)$ consisting of the  images $b'$ of
boundary states $b$ under $b\mapsto b'$. More explicitly,
\eqn\ggef{
{\cal G}\equiv \{ (b_0;b_1,\ldots,2b_n+1) \in P_+' \, : \,
2 \sum_{i=1}^{n} b_i \leq k \,\}\,.}
Then the right hand side of \twischargdiff\ becomes
\eqn\suoddkoddb{\eqalign{
\sum_b \N_{\lambda a}{}^b\, \hat{q}^\omega_{b} & =
\sum_{\ga'}b^\la{}_{\ga'}\,
\sum_{b'\in {\cal G}} \,
\Bigl[N'_{\ga' a'} {}^{b'} \, \dim_B(b')
-N'_{\ga' a'}{}^{J'b'} \, \dim_B(b')\Bigr] \cr
& = \sum_{\ga'}b^\la{}_{\ga'}\, \Bigl[
\sum_{b'\in {\cal G}} N'_{\ga' a'} {}^{b'} \, \dim_B(b')
- \sum_{J'b'\in {\cal G}} N'_{\ga' a'} {}^{b'} \, \dim_B(J'b') \Bigr]
\cr 
& = \sum_{\ga'}b^\la{}_{\ga'}\, \Bigl[
\sum_{b'\in {\cal G}} N'_{\ga' a'} {}^{b'} \, \dim_B(b')
+ \sum_{J'b'\in {\cal G}} N'_{\ga' a'} {}^{b'} \, \dim_B(b') \Bigr]
\qquad ({\rm mod}\ M_B)
\,,}}
where we have used in the final line that
$\dim(J'b') = - \dim(b')\, ({\rm mod}\ M_B)$, as is shown in
appendix~B. If $b'\in {\cal G}$, then $\hat{b}=J'b'\not\in {\cal G}$,
since $\hat{b}_1=k+1-b_1-2\sum_{i=2}^n b_i$, and thus 
\eqn\ggefa{
2 \sum_{i=1}^{n} \hat{b}_i = 2 k+2 - 2 \sum_{i=1}^{n} b_i
\geq k+2 \,.}
Conversely, if $b'\not\in {\cal G}$, then either $J'b'\in {\cal G}$, or
$J'b'=b'$ (\ie\ $b_0'=b_1'$). Thus $P_+'$ consists of the three
disjoint sets ${\cal G}$, $J'{\cal G}$, and the $J'$-fixed points 
$f=(f_1;f_1,f_2,\ldots,
f_n)$. Given that
$a'$ is a spinor weight while $\gamma'$ is not, $b'$ must be a spinor
weight in order for the fusion rule coefficient to be
non-trivial. Since $k$ is even, there are no spinor weights that are
fixed points of $J'$ (in particular, $f_n= k$ (mod 2) for fixed points
of $\widehat{B}_{n}$ at level $k+2$). Thus we can replace \suoddkoddb\
with 
\eqn\suoddkoddc{\eqalign{
\sum_b \N_{\lambda a}{}^b\, \hat{q}^\omega_{b} & = 
\sum_{\ga'}b^\la{}_{\ga'}\,
\sum_{b'\in P_+'} N'_{\ga' a'} {}^{b'} \, \dim_B(b') \qquad
({\rm mod}\ M_B) \cr
& =  \sum_{\ga'}b^\la{}_{\ga'}\, \dim_B(\ga') \, \dim_B(a')\, \qquad
({\rm mod}\ M_B) \cr
& = \dim_A(\lambda)\, \dim_B(a') \qquad
({\rm mod}\ M_B) \,.}}
Hence the ansatz \suoddodda\ solves \twischargdiff\ with
$M^\omega=M_B$, where $M=M_B$ is the untwisted charge number
corresponding to $\widehat{B}_n$ at
level $k+2$. Using conjecture B \refs{\bou} this equals
\eqn\mbfor{
M_B = {k+2n+1 \over {\rm gcd} \{ k+2n+1, L_B \}} \,, \qquad
L_B = {\rm lcm} \{1,2,\ldots,2n-1\} \,.}
Since $k$ is even, $k+2n+1$ is odd, and thus $M_B$ agrees, for
$n\geq 2$, with the expression for ${\rm su}(2n+1)$ at level $k$. It
follows that the ansatz \suoddodda\ solves \twischargdiff\ with
$M^\omega=M$. 

Finally, \suoddodda\ is equivalent to our proposed ansatz
\suoddp. Indeed, as is explained in appendix~A, we have the relation
\eqn\mir{
2^n\, {\rm dim}_{C}(a)={\rm dim}_{B}(a')\,.
}
Thus the charges \suoddodda\ differ by an overall factor $l=2^n$ from 
\suoddp. Since $k$ is even, $M^\omega$ is odd, and thus $2^n$ is
coprime to (and invertible modulo) $M^\omega$. Hence \suoddodda\ is
equivalent to
\suoddp; in particular, \suoddp\ therefore also satisfies
\twischargdiff\ with $M^\omega=M$.

\newsec{The analysis for ${\rm su}(2n)$}

 Next we turn to ${\rm su}(2n)$ with $\omega=C$, for which
the twisted algebra behaves differently from su$(2n+1)$. As was
explained in \refs{\bfs,\gg}, the $\omega$-twisted boundary states are
labelled by the level $k$ $\omega$-twisted weights of
$\g=\widehat{{\rm su}}(2n$), or alternatively the level $k$ weights of
the twisted Lie algebra $\g^\omega=A_{2n-1}^{(2)}$. They can be
equated with all ($n+1$)-tuples $(a_0;a_1,\ldots,a_n)$ where
$k=a_0+a_1+2a_2+\cdots+2a_n$ and $a_i\in \Zop_{\ge 0}$. Once again,
the invariant subalgebra $\bar{\g}^\omega$ is $C_n$, and the twisted
D-brane charge $q_a^\omega$ is given by \suoddp. 

In \refs{\gg} we remarked that the $\psi$-matrix can be interpreted as
a submatrix of $S$ (rescaled by $\sqrt{2}$) for $\widehat{B}_n$ level
$k+1$, using the identifications 
$\mu\mapsto \mu''=(\mu_0+\mu_1+1;\mu_1,\ldots,\mu_n)\in
P_+'$ (for $\omega$-invariant weights $\mu\in\E$) and
$a\mapsto a'=(a_0;a_1,\ldots,a_{n-1},2a_n+1)\in
P_+'$. [The $\omega$-invariant weights $\mu\in\E$ are characterised by
the condition $k\geq 2\mu_1+\cdots +2\mu_{n-1}+\mu_n$.]
As before we use primes to denote $\widehat{B}_n$ level $k+1$ 
quantities. In order to find an expression for the NIM-rep
coefficients in terms of $\widehat{B}_{n}$ fusions at $k'=k+1$, we
need to express the $\widehat{A}_{2n-1}$ ratios $S_{\la\mu}/S_{0\mu}$
in terms of the $\widehat{B}_{n}$ ratios
$S'_{\ga\mu''}/S'_{0\mu''}$. The usual way to do this involves
branching rules, but there is no embedding of so($2n+1$) into
su$(2n$). Fortunately, it is possible, nonetheless, to write any 
su$(2n$) character, restricted to $\omega$-invariant weights
$\mu\in{\cal E}$, in terms of so$(2n+1$) characters (this possibility
was missed in \refs{\gg}). This is done by comparing the embeddings
sp$(2n)\subset {\rm su}(2n)$ and sp$(2n)\subset{\rm so}(2n+1)$. What
we find is that, for any $1\le m< n$ and any $\mu$,  
\eqn\susoonee{
{\rm ch}_{\Lambda_m}[\mu]=
{\rm ch}_{\Lambda_{2n+1-m}}[\mu]
=\sum_{i=0}^m (-1)^i \,
{\rm ch}'_{\Lambda_{m-i}'}[\mu'']\,,}
and
\eqn\susotwo{{\rm ch}_{\Lambda_n}[\mu]
={\rm ch}'_{2\Lambda_n'}[\mu'']
+\sum_{i=0}^{n-1}
(-1)^i\, {\rm ch}'_{\Lambda_{n-i}'}[\mu'']\,,}
where we have defined
$$
{\rm ch}_{\lambda} [\mu] \equiv
{\rm ch}_\lambda \left(-2\pi i (\mu+\rho)/(k+2n) \right)={S_{\la\mu}\over 
S_{0\mu}} \,,
$$
and similarly for ch$'_{\la'}[\mu'']$.
Here, the $\Lambda_i$ and $\Lambda_i'$ are the fundamental weights for
$A_{2n-1}$ and $B_{n}$, respectively. More generally, we can
write  
$${\rm ch}_\la[\mu]=\sum_\gamma
b^\la{}_\gamma\,
{\rm ch}'_\gamma[\mu'']\,,$$
where these branching rule-like coefficients
$b^\la{}_\gamma$ are necessarily integers but can be 
negative.\footnote{$^\ddagger$}{Virtual branching rules of this kind
were considered before in \refs{\pss} where they were called
`subjoinings'; we thank Mark Walton for bringing this reference to our
attention.} 
Note that all $\gamma$ appearing in these decompositions will be
non-spinors (since this is true of the fundamental weights
$\Lambda_m$). 

We can now compute
$${\cal N}_{\la a}{}^b=\sum_{\mu\in{\cal E}} \psi_{a\mu}
{S_{\la\mu}\over S_{0\mu}} \,
\psi_{b\mu}= 2\sum_{\mu''\in{\cal G}'}\sum_\gamma
S'_{a'\mu''}\, b^\la{}_\gamma  \;
{S'_{\gamma \mu''}\over S'_{0\mu''}}\,
S'_{b'\mu''}\,,$$
where the set ${\cal G}'$ consists of all images 
$\mu\mapsto \mu''\in P_+'$, \ie\ of all $\mu''\in P_+'$ for which
$\mu_0''>\mu_1''$. We need to extend the sum over $\mu''$ to
all of $P_+^{k+1}(B_n)$. For this, we observe that since 
${\cal G}'$ consists of all $\mu''$ with $\mu_0''>\mu_1''$, any 
weight $\nu\in P_+'$ is either exclusively in ${\cal G}$, or $J'\nu$
is, or $\nu=J'\nu$ is a fixed point. Because $a'$ 
is a spinor, if $\nu=J'\nu$ then
$S'_{a'\nu}=S'_{a',J'\nu}=-S'_{a'\nu}$ and so $S'_{a'\nu}=0$.
Because $a'$ and $b'$ are spinors and $\gamma$ is not,
$S'_{a',J'\nu} {S'_{\gamma, J'\nu}}S'_{b',J'\nu}/S'_{0,J'\nu}=S'_{a'\nu}
{S'_{\gamma \nu}}S'_{b'\nu}/S'_{0\nu}$. Thus we obtain
\eqn\suevnim{{\cal N}_{\la a}{}^b=\sum_\gamma b^\la{}_\gamma
N'_{\gamma a'}{}^{b'}\,,}
where $N'_{a'\gamma}{}^{b'}$ is the $\widehat{B}_{n}$ fusion rule at
$k'=k+1$. In particular, we obtain from \susoonee\ the nice
formula 
\eqn\suevfund{{\cal N}_{\Lambda_m a}{}^b=\sum_{i=0}^m (-1)^i\,
N'_{\Lambda'_{m-i}a'}{}^{ b'}\,,}
for any $1\le m<n$, with the obvious minor modification for
$m=n$. Incidentally, the NIM-rep for the simple current $J^i$ is
trivial for $i$ even, and for $i$ odd corresponds to the interchange
$a_0\leftrightarrow a_1$.  

\noindent As in the previous case we now make the ansatz (see 
\suoddodda) $\hat{q}^\omega_a={\rm dim}_B(a')$. In order to verify 
that this obeys \twischargdiff, we observe
\eqn\suevcal{\eqalign{
{\rm dim}_A(\la)\,{\rm dim}_B(a')
=&\sum_\gamma b^\la{}_\gamma \, {\rm dim}_B(\gamma)\,
{\rm dim}_B(a')\cr
=&\sum_\gamma\sum_b b^\la{}_\gamma N'_{\gamma a'}{}^{b'}\,
{\rm dim}_B(b')
=\sum_b{\cal N}_{\la a}{}^b\, {\rm dim}_B(b')\qquad ({\rm mod}\
2^nM_B)\,,}} 
using Conjecture B$^{spin}$. If conjecture B holds, then
$M_B$ equals manifestly $M_A$ (compare \Mdef\ with \yar). Also, we
know dim$_B(a')=2^n{\rm dim}_C(a)$, from appendix~B. Thus  we can
divide \suevcal\ by $2^n$ and the desired result follows.  

As is explained in appendix~D, we can {\it prove} Conjecture
B$^{spin}$ (assuming Conjecture B) provided $4$ does not divide
$M_B$. For example this is automatic provided $k$ and $-2n$ are not
congruent mod 8. In particular, this covers therefore the case when
$k$ is odd.  

Finally, let us sketch why the NIM-rep coefficient $\N_{\lambda a}^b$
does indeed equal the corresponding tensor product coefficient
$T_{\lambda a}{}^b$ for the invariant subalgebra
$\overline{\g}^\omega=C_n$, for sufficiently large level. One way to
see this uses the expression, derived in \refs{\pz}, for 
$\N$ in terms of fusions $\widehat{N}$ for sp($2n)$ at level 
$k+n-1$:\footnote{$^\star$}{The fact that a given NIM-rep may be
related to different fusion rules was first observed in
\refs{\quella}.}  
\eqn\foursix{
\N_{\lambda a}{}^b=\sum_{l=0}^{n-1}\sum_{n\ge i_1>\cdots>i_l>1}
(-1)^{\lfloor{i_1\over 2}\rfloor+\cdots+\lfloor{i_l\over 2}\rfloor}
\sum_cb^\lambda{}_c\widehat{N}_{ac}{}^{\sigma_{i_1}\cdots\sigma_{i_l}
\sigma_1^{i_1+\cdots+i_l}(b)}\,,}
where the $b^\lambda{}_c$ are the sp$(2n)\subset{\rm su}(2n)$
branching coefficients, and where 
$$\sigma_i[a_1,\ldots,a_n]=[a_{i-1},\ldots,a_1,
k+2n-\sum_{j=1}^ia_j-2\sum_{j=i+1}^na_j,a_{i+1},\ldots,a_n]\,.$$
For $k$ much larger than $\sum_{i=1}^n(\lambda_i+a_i)$, it is easy to
see that for any choice of $n\ge i_1>\cdots>i_l>1$ (for $l>0$) some
component of
$\sigma_{i_1}\cdots\sigma_{i_l}\sigma_1^{i_1+\cdots+i_l}(b)$ will be
large --- \eg, if $i_1+\cdots+i_1$ is odd it will be the
$(i_1-i_2+\cdots\pm i_l \mp 1)$-th one. For this reason, all fusion
coefficients $\widehat{N}$ in \foursix\ will vanish except the $l=0$
one, and we obtain $\N_{\lambda a}{}^b=
\sum_cb^\lambda{}_c\widehat{N}_{ac}{}^b={T}_{\lambda a}{}^b$.

\newsec{The analysis for ${\rm so}(2n)$}

The final algebras we will discuss are ${\rm so}(2n)=D_n$
with the automorphism $\omega$ corresponding to chirality-flip (\ie\ 
$\omega$ interchanges the Dynkin labels
$\la_{n-1}\leftrightarrow\la_n$). The boundary states are all
$n$-tuples $(a_0;a_1,\ldots,a_{n-1})$ for which 
$k=a_0+2a_1+2a_2+\cdots+2a_{n-2}+a_{n-1}$. Let primes denote the
$\widehat{B}_{n-1}$ level $k+1$ quantities as usual.
The NIM-rep coefficients can be interpreted \refs{\gg}\ in terms of
fusions $N'$, and branching coefficients
$b^\la{}_{\ga'}$ of $B_{n-1}\subset D_n$,
\eqn\nimso{
\N_{\la a}{}^b=\sum_{\ga'}
b^\la{}_{\ga'}\,\bigl(N'_{\ga' a'}{}^{b'}
-N'_{\ga' a'}{}^{J'b'}\bigr)\,,}
where $a'=(a_0+a_1+1;a_1,\ldots,a_{n-1})\in P_+'$. In particular,
\eqn\soso{\N_{\Lambda_i a}{}^b=N'_{\Lambda_i'
a'}{}^{b'}+N'_{\Lambda_{i-1}'a'}
{}^{b'}\,,}
for $i<n-1$ (where the term $N'_{\Lambda_0'a'}{}^{b'}$ is defined to be
$\delta_a{}^b$), as well as
\eqn\sosotwo{\N_{\Lambda_na}{}^b=\N_{\Lambda_{n-1}a}{}^b
=N'_{\Lambda_{n-1}'a'}{}^{b'}\,.}
The simple current matrix $\N_{J_v}$ is trivial, while the simple
current matrices $\N_{J_s}$ and $\N_{J_c}$ are equal and correspond to
the symmetry of the $D_n^{(2)}$ Dynkin diagram, \ie\ they send $a$ to 
$(a_n;a_{n-1}, \ldots,a_0)$.

The invariant subalgebra $\bar{\g}^\omega$ here is $B_{n-1}$. 
Our ansatz for the charge is
\eqn\soans{
q^\omega_a = \dim_B([a_1,\ldots,a_{n-1}]) = \dim_B (a') \,.}
The right hand side of \twischargdiff\ is then
\eqn\soa{
\sum_b {\cal N}_{\la a}{}^{b} q^\omega_b = \sum_{\ga'}
b^\la{}_{\ga'}\, \Bigl[
\sum_{b'\in{\cal G}'} N'_{\ga' a'}{}^{b'}\, \dim_B (b')
- \sum_{b'\in{\cal G}'} N'_{\ga' a'}{}^{J'b'}\, \dim_B (b') \Bigr] \,,}
where ${\cal G}'$ is the subset of $P_+'$ that appeared already in
section~4. As explained there, any 
$b\in P_+'$ is either in ${\cal G}'$, or $J'b$ is in
${\cal G}'$, or $J'b=b$. Using as before the fact that
$\dim_B(J'b)=-\dim_B(b)$ modulo $M_B$, we therefore have that
\eqn\sob{\eqalign{
\sum_b {\cal N}_{\la a}{}^{b} q^\omega_b & = \sum_{\ga'}
b^\la{}_{\ga'}\, \Bigl[
\sum_{b\in P_+'} N'_{\ga' a'}{}^{b}\, \dim_B (b)
- \sum_{J'b=b} N'_{\ga' a'}{}^{b} \, \dim_B (b) \Bigr] 
\quad ({\rm mod}\ M_B) \cr
& = \dim_D(\lambda) \, q^\omega_{a}
- \sum_{\ga'} b^\la{}_{\ga'}\, \sum_{J'b=b} N'_{\ga' a'}{}^{b} \,
\dim_B (b) \qquad ({\rm mod}\ M_B) \,.}}
Using conjecture B and D, the explicit formulae \Mdef\ imply that
$M_B$ for $\widehat{B}_{n-1}$ at level $k+1$ equals $M_D$ for
$\widehat{D}_n$ at level $k$, and thus the identity holds modulo
$M_D$. Thus if we can ignore the last term, the ansatz \soans\ solves
\twischargdiff\ with $M^\omega=M_D$. 

It therefore only remains to analyse the last term in \sob. Since
$\dim_B(J'b)=-\dim_B(b)$ modulo $M_B$, if $b$ is a fixed point of $J'$
we have $2\dim_B(b)=0$ modulo $M_B$. If $M_B$ is odd, as is the case
for example when $k$ is odd, then $\dim_B(b)=0$ modulo $M_B$, and we can
ignore the last term in \sob. This leaves us with the case where both $k$
and $M_B$ are even. It is easy to see that $J'b=b$
implies that $b_{n-1}=k+1$ $({\rm mod}\ 2)$. Thus for $k$ even the
fixed point is necessarily a spinor of $B_{n-1}$, so $2^n$ must divide
its dimension (see the argument given in detail in the next
section). Provided $2^{n+1}$ does not divide $M_B$, the facts that
$M_B/2$ and $2^n$ both divide dim$_B(b)$ implies that $M_B$ itself
must divide dim$_B(b)$, and thus we can again ignore the last term in
\sob. 

This leaves with the case when $2^{n+1}$ does divide $M_B$. In this
case we use Conjecture B$^{spin}$ to deduce that $2^nM_B$ divides  
$({\rm dim}(J'0)-1) {\rm dim}_B(b)$. Together with the fact (proved in
appendix~B) that dim$_B(J'0)=-1+\ell M_B$ for some $\ell\in\Zop$, we
can actually deduce the stronger result that
$2^{n-1}M_B$ divides dim$_B(b)$. Thus we can again ignore the last term in
\sob, and we are done.

It is clear that, for large $k$, the NIM-rep coefficient 
$\N_{\la a}{}^b$ in \nimso\ becomes the $\bar{\g}^\omega=B_n$ tensor
product coefficient $T'_{\la a}{}^{b}$:  
for large $k$, the term $N'_{\ga'a'}{}^{J'b'}$ vanishes.

\newsec{Uniqueness }

In this section we prove statement (ii) of section 1.2. Consider
first the case of su$(2n)$. Choose an integer $q_{b'}'$ for each
spinor $b'\in P_+^{k+1} (B_n)$, and an integer $m'$ such that 
\eqn\sixone{
{\rm dim}_A(\lambda)\,q_{b'}'2^n
=\sum_{\gamma,c'}b^\lambda{}_\gamma N'_{b'\gamma}
{}^{c'}\,q'_{c'}2^n \qquad ({\rm mod}\ 2^nm')\,,}
for all $\lambda\in P_+^k(A_n)$ and all spinors $b'\in P_+^{k+1}(B_n)$
(the multiplication by $2^n$ is for later convenience). We require
gcd$\,q'_{b'}=1$. Then $2^nm'$ divides $2^nM_A$ (by the argument in
section 1.2), and so must also divide $m_B^{spin}$ (by Conjecture
$B^{spin}$).  

It is a classical result (see \eg\ \refs{\brb}) that any $B_n$
character ch$'_{b'}$ can be expressed as a polynomial over the
integers $\Zop$ in the fundamental characters ch$'_{\Lambda_i'}$
($i\le n$). If the weight $b'$ is a spinor, then every term in this
polynomial will contain an odd number of ch$'_{\Lambda_n'}$'s. 
But
\eqn\sixtwo{
({\rm ch}'_{\Lambda_n'})^2=1+\sum_{i=1}^{n-1}{\rm ch}'_{\Lambda'_i}
+{\rm ch}'_{2\Lambda'_n}\,,}
and so the character ch$'_{b'}$ of any $B_n$ spinor can be expressed
as the product of ch$'_{\Lambda'_n}$ with some polynomial in
ch$'_{\Lambda'_i}$ ($i<n$) and ch$'_{2\Lambda'_n}$, or more precisely
some combination 
ch$'_{b'-\Lambda'_n} +\sum_j\ell'_j {\rm ch}'_{\gamma^{(j)}}$ of
non-spinors (for $\ell_j'\in\Zop$). By construction, each
$\gamma^{(j)}+\Lambda'_n\prec b'$, \ie\ $b'-\gamma^{(j)}-\Lambda'_n$
is a nonzero sum of positive roots. Now, from \susoonee, \susotwo\ we
obtain 
ch$'_{\Lambda'_i} ={\rm ch}_{\Lambda_i}-{\rm ch}_{\Lambda_{i-1}}$, for 
$i<n$, and 
ch$'_{2\Lambda_n'} ={\rm ch}_{\Lambda_n}-{\rm ch}_{\Lambda_{n-1}}$. 
Thus the character ch$'_{b'}$ of any $B_n$ spinor can be expressed as
the product of ch$'_{\Lambda'_n}$ with some combination
$\sum_i\ell_i{\rm ch}_{\lambda^{(i)}}$ (with integer coefficients
$\ell_i$) of characters of $A_n$ weights $\lambda^{(i)}$. 

We are interested here in Weyl dimensions, \ie\ in the evaluation of
these characters at 0. In particular we obtain
\eqn\sixthree{\eqalign{
{\rm dim}_B(b')q'_{\Lambda'_n} & =\sum_i\ell_i\,
{\rm dim}_A(\lambda^{(i)}) \,2^n\, q'_{\Lambda'_n} \cr
& =\sum_{c'}\bigl(N'_{b'-\Lambda'_n,\Lambda'_n}{}^{c'}
+\sum_j \ell_j'\,N'_{\gamma^{(j)}\Lambda'_n}{}^{c'}\bigr)q'_{c'}
2^n\qquad({\rm mod}\ 2^nm')\,,}}
where $c'\prec b'$ are spinors. Inductively, we get 
$q'_{b'}2^n={\rm dim}_B (b')q'_{\Lambda_n'}$ (mod $2^nm'$). Hence
$1={\rm gcd} q'_{b'}={\rm gcd}\{ q'_{\Lambda_n'},m'\}$, and we get
statement (ii) of section 1.2. 

Thus there is a unique solution to
\twischargdiff\ for su$(2n$), up to equivalence. Note that uniqueness
here follows for much the same reason uniqueness  occurs for the
untwisted case: there is a boundary condition here (namely
$a=[0,\ldots,0]$) which acts as an identity. Uniqueness for su$(2n+1)$
at odd  level follows immediately from the $\widehat{C}_{n}$ level
$(k-1)/2$ analysis of \refs{\bou}. The uniqueness argument for
su($2n+1)$ when $k$ is even is similar to that given above for 
su$(2n)$: by those arguments we get 
$q'_a={\rm dim}_B(a')=-{\rm dim}_B (J'a')$ (mod $m'$) for boundary
states $a\in {\cal B}$, and uniqueness follows. The argument for
so$(2n)$ is the same as that of su$(2n+1)$ at even level.

\newsec{Conclusion}

In this paper we have shown (under the assumptions detailed in 1.2)
that the charge group of the $\omega$-twisted D-branes of WZW models
agrees with that of the untwisted D-branes. This is in nice agreement
with the recent K-theory calculation \refs{\braun}. 

We have also shown that the charge of the $\omega$-twisted D-brane
corresponding to the twisted representation $a$ has a simple
interpretation: it is the dimension of the conformal highest weight
space of the representation $a$. As for the case of an untwisted
D-brane, the charge associated to the D-brane has therefore a simple
string theoretic interpretation: it is the multiplicity of the ground
state in the open string between the fundamental D0-brane (the
untwisted D-brane corresponding to the identity representation) and
the brane in question. [This follows simply from the fact that the
states of the open string between the fundamental D0-brane and any
D-brane labelled by $x$ (where $x$ is either an untwisted
representation $\mu$ or a twisted representation $a$) transforms in
the irreducible representation $x$.] In the full supersymmetric
formulation of the WZW models, this number therefore has an
interpretation as an intersection number. This observation may help to
give a more conceptual argument for why the twisted and the untwisted
charge groups should coincide. 

In this paper we have restricted attention to the case of the
classical algebras with $\omega$ of order two. We expect that the
remaining cases (namely E6 with $\omega$ being charge conjugation, and
SO$(8)$ with $\omega$ being triality) should work out similarly.
Our arguments relied on a number of intriguing Lie theoretic
properties. It would be nice to find more conceptual arguments for
them.

We have also found simple formulae for the NIM-rep coefficients for
the fundamental representations. These NIM-rep coefficients are
truncations of the tensor product coefficients of the invariant
subalgebra. This gives strong support to the suggestion that these
NIM-reps are in fact the twisted fusion rules.

\vskip 1cm

\centerline{{\bf Acknowledgements}}\par\noindent

\noindent We thank Ilka Brunner, Thomas Quella and Mark Walton for
useful conversations and correspondences, and in particular Stefan
Fredenhagen for communicating to us his argument that $M^\omega$
always has to be a factor of $M$. This work was done while the authors
were visiting BIRS; we are very grateful for the wonderful working 
environment we experienced there! The research of TG is also supported
in part by NSERC.

\vskip1.5cm

\appendix{A}{The proof of the miraculous dimension formula}

 The purpose of this subsection is to prove, for any $n\ge 2$
and any
highest weight $\la=[\la_1,\ldots,\la_n]$ of ${\rm sp}(2n)=C_n$, the
identity
\eqn\MirDim{
2^n\, {\rm dim}_{C}(\la)={\rm dim}_{B}(\la')\,,}
relating the Weyl dimension of the $C_n$ weight $\lambda$ to that of
the $B_n$ weight $\lambda'=[\la_1,\ldots,\la_{n-1},2\la_n+1]$. For
notational
clarity, in this appendix we shall let primes denote $B_n$ quantities
(or sometimes $\widehat{B}_n$ level $k$ quantities), while $C_n$ (or
$\widehat{C}_n$ level $k+n-2$) quantities are unprimed.

The denominator identity of the simple Lie algebras yields a
useful expression for the quantum dimensions. For $\widehat{B}_{n}$
level $k$, let $\kappa= k+2n-1$ and write  
$\mu[i]'=\sum_{j=i}^{n-1}(\mu_j+1)+(\mu_n+1)/2$ as usual. Then
\eqn\Bndim{
{S'_{\mu 0}\over S'_{00}}=
\prod_{1\le i<j\le n}{\sin(\pi(\mu[i]'-\mu[j]')/\kappa)
\over\sin(\pi(0[i]'-0[j]')/\kappa)}
{\sin(\pi(\mu[i]'+\mu[j]')/\kappa)
\over\sin(\pi(0[i]'+0[j]')/\kappa)} 
\prod_{i=1 }^n{\sin(\pi\,\mu[i]'/\kappa)
\over\sin(\pi\,0[i]'/\kappa)}\,.}
For $\widehat{C}_{n}$ level $k+n-2$, let 
$\mu[i]=\sum_{j=i}^{n}(\mu_j+1)$ as usual. Then
\eqn\Cndim{
{S_{\mu 0}\over S_{0 0}}=\prod_{1\le i<j\le n}
{\sin(\pi(\mu[i]-\mu[j])/(2\kappa))
\over\sin(\pi(0[i]-0[j])/(2\kappa))}
{\sin(\pi(\mu[i]+\mu[j])/(2\kappa))
\over\sin(\pi(0[i]+0[j])/(2\kappa))}
\prod_{i=1}^n{\sin(\pi\,\mu[i]/\kappa)\over
\sin(\pi\,0[i]/\kappa)}\,.}
The Weyl dimension formula is obtained from these by taking the
$k\rightarrow\infty$  limit, and using the
asymptotic formula sin$(x)\approx x$ for $x$ small.

Note that each $\la'[i]'=\la[i]$. Thus, the ratio of the
$B_{n}$ quantum dimension of $\la'$ with the $B_{n}$ quantum
dimension of the spinor $0'=\Lambda_n$, is given by the formula \Cndim\
(for the same value of $\kappa$),  except that the `$2\kappa$'s
there are replaced with `$\kappa$'s. Taking the $k\rightarrow\infty$
limit, and noting that the dimension of the $B_n$ spinor
$\Lambda_n$ is $2^n$, we get the desired \MirDim.

It would be interesting to obtain a more conceptual, algebraic proof of
\MirDim, as opposed to our direct calculation from the Weyl dimension
formula. Two observations along these lines can be made. The first is
that the source of the $2^n$ factor is presumably the $B_n$
fundamental spinor. The second is that this relation \MirDim\ clearly
does not generalise naturally to quantum dimensions, and thus cannot
be obtained merely by playing with tensor products or character
manipulations. It is for making this second point that we wrote down
the preceding {\it quantum} dimension formulae, rather than the simpler
{\it Weyl}
dimension formulae.

\appendix{B}{Dimensions of simple currents}

Let $J$ be the simple current of $\widehat{B}_{n}$ level $k'$.
So $J[a_1,\ldots,a_n]=
[k'-a_1-2a_2-\cdots-2a_{n-1}-a_n,a_2,a_3,\ldots,a_n]$.
Put $\kappa=k'+2n-1$, and $M_B=\kappa/{\rm gcd}\{L_B,\kappa\}$ as usual,
where $L_B$ is as in \Mdef.
\medskip

\noindent{{\bf Claim.}} Let $\nu$ be any weight of $B_{n}$ level $k'$.
Then the dimensions of $J\nu$ and $\nu$ are related by
\eqn\scdim{
{\rm dim}_B(J\nu)= -{\rm dim}_B(\nu)\qquad({\rm mod}\ M_B)\,.}

To prove this, first note that it is sufficient to show that the
dimension dim$_B(J0)={\rm dim}_B[k',0,\ldots,0]$ is congruent to $-1$
(mod $M_B$).  From \Bndim\ we obtain
\eqn\dimsc{
{\rm dim}_B(J0)=\left({\kappa\atop 2n-1}\right)
+\left({\kappa-1 \atop 2n-1}\right)\,.}

Let $p$ be any prime dividing $M_B$. Let $p^\alpha$ and $p^\beta$ be the
exact powers of $p$ dividing $\kappa$ and $L_B$, respectively (we say
$p^a$ {\it exactly divides} $N$ if $N/p^a$ is coprime to $p$). So we
know that the largest power of $p$ which divides any number
$1\le\ell\le 2n-1$, is $p^\beta$. We also know
$\alpha> \beta\ge 0$ and $p^{\alpha-\beta}$ exactly divides $M_B$.

Prime factorisations apply to rational
numbers $r$ as for integers. Once again we get a unique factorisation
$r=\pm\prod_i p_i^{a_i}$, where now the exponents
${\rm exp}_{p_i}(r)\equiv a_i\in\Zop$ can be negative. We say that $p^a$
{\it divides} $r$ if $a\le {\rm exp}_p(r)$, and we say that $p$ is
{\it coprime to} $r$, if exp$_p(r)=0$. For example, exp$_2(.75)=-2$,
and 5 is coprime to $.75$.

We will show that $p^{\alpha-\beta}$ divides
$$
\left({\kappa\atop 2n-1}\right)={\kappa\over 2n-1}
\prod_{\ell=1}^{2n-2}{\kappa-\ell\over\ell}\,,
$$
and that
$$
\left({\kappa-1\atop 2n-1}\right)=
\prod_{\ell=1}^{2n-1}{\kappa-\ell\over\ell}
$$
is congruent to $-1$ (mod $M_B$). To see this, take any
$1\le \ell\le 2n-1$ and let $i={\rm exp}_p(\ell)$. Then
$\gamma\le\beta<\alpha$ and $p$ is coprime to the rational number
$(\kappa-\ell)/\ell=((\kappa/p^i)-(\ell/p^i))/ (\ell/p^i)$. This
immediately gives us the first statement, because $p^{\alpha -\beta}$
clearly divides the first fraction. To get the second statement, note
that $\ell/p^i$ is coprime to $p$ and hence is invertible mod
$p^{\alpha-\beta}$, and so we obtain
$$
(\kappa-\ell)/\ell=
((\kappa/p^i)-(\ell/p^i))(\ell/p^i)^{-1}
=
(0-(\ell/p^i))(\ell/p^i)^{-1}
=-1\qquad \left({\rm mod}\ p^{\alpha-\beta}\right)\,.
$$

\appendix{C}{The example of ${\rm su}(7)$}

In this appendix we describe the NIM-rep for ${\rm su}(7)$,
level $k=1,2,3$. Let us recall that the boundary states are labelled
by triples of integers $a=[a_1,a_2,a_3]$, where 
$k\geq 2a_1+2a_2+2a_3$. These triples can be identified with a subset
of the $C_3$ weights at $k_C=k+3$. To each such weight, associate the
$B_3$ weight $a'=[a_1,a_2,2a_3+1]$; this defines a dominant highest
weight for $B_3$ at level $k_B=k+2$. Let us denote the first
fundamental representation of ${\rm su}(7)$ by ${\bf 7}$, and likewise
for ${\rm so}(7)$. Then our NIM-rep formula is simply
\eqn\suseven{
{\cal N}_{{\bf 7} a}{}^b = N'_{{\bf 7} a'}{}^{b'} -
N'_{{\bf 7} a'}{}^{J'b'} \,,}
where $N'_{{\bf 7} a'}{}^{b'}$ are the fusion rules of ${\rm so}(7)$
at level $k_B=k+2$. In this manner one easily calculates
\eqn\susevenr{\eqalign{
{\bf 7} \otimes [0,0,0]_{1} & = [0,0,0]_{1} \oplus [1,0,0]_{2}
\cr
{\bf 7} \otimes [1,0,0]_{2} & = [0,0,0]_{2} \oplus
[1,0,0]_{2} \oplus [0,1,0]_{2} \oplus [2,0,0]_{4}\cr
{\bf 7} \otimes [0,1,0]_{2} & = [1,0,0]_{2} \oplus
[0,0,1]_{2} \oplus [0,1,0]_{3} \oplus [1,1,0]_{4} \cr
{\bf 7} \otimes [0,0,1]_{2} & = [0,1,0]_{2} \oplus [0,0,1]_{3}
\oplus [1,0,1]_{4} \,, }}
where $[a_1,a_2,a_2]$ label the boundary states (in $C_3$ notation),
and the index  denotes the value of the level $k$ for which the
relevant representation appears first in the NIM-rep. These twisted
fusion rules are indeed a truncation of the $C_3$ tensor product
coefficients: under the embedding of $C_3$ into su$(7)$, the ${\bf 7}$
of su$(7)$ becomes $[0,0,0]\oplus [1,0,0]$. Equation \susevenr\ should
therefore be compared with
\eqn\susevens{\eqalign{
\Bigl([0,0,0]\oplus [1,0,0]\Bigr) \otimes [0,0,0] & =
[0,0,0]\oplus [1,0,0] \cr
\Bigl([0,0,0]\oplus [1,0,0]\Bigr) \otimes [1,0,0] & =
[0,0,0] \oplus
[1,0,0] \oplus [0,1,0] \oplus [2,0,0]\cr
\Bigl([0,0,0]\oplus [1,0,0]\Bigr) \otimes [0,1,0] &
= [1,0,0] \oplus
[0,0,1] \oplus [0,1,0] \oplus [1,1,0] \cr
\Bigl([0,0,0]\oplus [1,0,0]\Bigr) \otimes [0,0,1] & =
[0,1,0] \oplus
[0,0,1] \oplus [1,0,1] \,. }}
For large $k$, these two expressions agree indeed.

\appendix{D}{Verification of Conjecture B$^{spin}$ for most levels}

We know \refs{\gep}\ that the fusion ring of $\widehat{B}_n$ level
$k'$, for example, is the quotient of the character ring of $B_n$, say,
by the fusion ideal ${\cal I}_{k'}$, which can be thought of as all 
linear combinations (with coefficients in $\Zop$) in the $B_n$  
characters ch$_\lambda$, which vanish at all values 
$-2\pi i(\mu+\rho)/(k'+2n-1)$ for $\mu\in P_+^{k'}(B_n)$. We can split
${\cal I}_{k'}$ into a sum ${\cal I}_n+ {\cal I}_s$ of spinor and
non-spinor contributions. By dim$_B(\sum\ell_i{\rm ch}_{b^{(i)}})$ we
mean $\sum \ell_i{\rm dim}_B(b^{(i)})$ (or equivalently, the
specialisation of that character sum to 0). Let $M_n$ (resp.\ $M_s$)
be the gcd of all dim$_B(\sum\ell_i{\rm ch}_{b^{(i)}})$, as one runs
over all elements of ${\cal I}_n$  (resp.\ ${\cal I}_s$). Then
$m_B={\rm gcd}\{M_n,M_s\}$ and $m_B^{spin}=M_s$. 

Now, ${\cal I}_{k'}$ is an ideal, so ch$_{\Lambda_n} {\cal I}_{k'}$ is
contained in ${\cal I}$, and hence 
ch$_{\Lambda_n} {\cal I}_n\subset {\cal I}_s$ 
and ch$_{\Lambda_n} {\cal I}_s\subset {\cal I}_n$. Thus $M_s$ divides
$2^nM_n$ and $M_n$ divides $2^nM_s$. In particular, $m_B,M_s,$ and
$M_n$ differ from one another only by powers of 2. 

As we know (see section 6), $2^n$ must divide the dimension of any
spinor, and hence $2^n$ must divide $M_s$. Thus as long as $2^n$ does
not divide $m_B$ (\eg\ if $m_B$ is odd), $m_B$ must equal $M_n$. 

Consider first $m_B$ odd. Then we know $2^n$ divides $M_s$ and $M_s$
divides $2^nM_n$. Thus $M_s=2^nM_n$, in agreement with conjecture
B$^{spin}$. 

Now suppose $m_B$ is even (so the level $k'$ is necessarily odd). From
the reasoning of section 6, we find that every element of ${\cal I}_s$
is of the form ch$_{\Lambda_n}\sum\ell_j{\rm ch}_{b^{(j)}}$ where each
$b^{(j)}$ is a non-spinor. It can be shown that, for 
$\mu\in P_+^{k'}(B_n)$, ch$_{\Lambda_n}[\mu] =0$ iff $\mu$ is a fixed
point of $J$. (Proof: If $J \mu=\mu$, then 
$S_{\Lambda_n \mu}=S_{\Lambda_n,J\mu}=-S_{\Lambda_n\mu}$. So
$S_{\Lambda_n\mu}=0$. Conversely, if $S_{\Lambda_n\mu}=0$, then  
$$
{S_{\Lambda_i\mu}\over S_{0\mu}}={S_{\Lambda_i,J\mu}\over S_{0,J\mu}}
\qquad  \forall i=1,\ldots,n\,,$$ 
and so $\mu=J\mu$.) 

Thus ch$_{\Lambda_n}\sum\ell_j{\rm ch}_{b^{(j)}}\in
{\cal I}_s$ iff $\sum\ell_j{\rm ch}_{b^{(j)}}[\mu]=0$ for all non-fixed
points $\mu\in P_+^{k'}(B_n)$. In particular, this means 
$\sum\ell_j{\rm ch}_{b^{(j)}}[\mu]=0$ for all non-spinors 
$\mu\in P_+^{k'}$. Thus 
$\sum\ell_j{\rm ch}_{b^{(j)}}[\mu]= -\sum\ell_j{\rm ch}_{Jb^{(j)}}[\mu]$ 
for all $\mu\in P_+^{k'}$. This means that $\sum\ell_j{\rm ch}_{b^{(j)}}$
is a sum of terms of the form ch$_b-{\rm ch}_{Jb}$. Since 
dim$_B({\rm ch}_\nu-{\rm ch}_{J\nu})=2\nu$ (mod $m_B$) and $m_B$ is
even, we get that dim$_B(\sum\ell_j{\rm ch}_{b^{(j)}})$ must also be
even. 

Thus $2^{n+1}$ must divide $M_s$. If 4 does not divide $m_B$, then
$M_s$ dividing $2^nm_B$ requires $M_s=2^nm_B$, in agreement with
conjecture B$^{spin}$. 

Note that, since $L_B$ is always even, 4 can divide $M_B$ only if 8
divides $k'+2n-1$. 

\listrefs

\bye